# Traceability Decentralization in Supply Chain Management Using Blockchain Technologies

Thomas Sermpinis[1], Christos Sermpinis[2]

[1]Department of Informatics, Aristotle University of Thessaloniki, Greece
[2]Department of Logistics, TEI of Central Macedonia, Greece

sermpinis@csd.auth.gr[1]. sermpinischris@protonmail.ch[2]

**Abstract**

With the increase of web users and applications with real time requests, the ability to identify, track and trace elements of a product as it moves in the supply chain is deemed necessary, and for many industries is even mandated by national or international regulations. Traceability presupposes the integrity and transparency of data that is saved and shared. This is a problem for current technologies, as there are many examples with tampered data and database vulnerabilities that resulted in serious implications and data loss. A solution to this problem can be the decentralization of the system, which will remove the central point of failure. To that effect, blockchain or DLT technologies, an emergent technology that enables the decentralization of a network can be used, by implementing a trustless model to achieve it. Blockchains are tamperproof and transparent, which means that by exploiting blockchain characteristics, traceability can be improved. A model that describes the decentralization process of the supply chain traceability part has been developed for this paper and is later evaluated and compared with the traditional system.

## 1. Introduction

Supply chain management is responsible for the management of the flow of millions of products and services every day, and traceability is one of the most important aspects of it. In accordance with ISO[1], traceability is the ability to trace the history, application or location of an entity, by means of recorded identifications (International Organization for Standardization, 1994). Product traceability can be distinguished in two types:

---

[1] The International Organization for Standardization

- Forward traceability, which is the ability to find locality of products from one or several given criteria, at every point of the supply chain (Jansen-Vullers et al., 2003)
- Backward traceability, which is the ability to find origin and characteristics of a product from one or several given criteria, at every point of the supply chain.

Jansen-Vullers et al. (2003) also suggests the four elements of traceability (Jansen-Vullers et al., 2003):

1. Physical lot integrity, which determines the traceability resolution;
2. Collection of tracing and process data;
3. Product identification and process linking; and
4. Reporting/system data retrieval.

Based on the above elements they provide a reference data model, which enables traceability information modelling and covers the requirements that their study produced.

The problem with traceability industry applications arises with the centralization of these systems which introduces a central point of failure. The use of databases and other centralized technologies, to store and manage traceability data that users will be able access, results in certain drawbacks since it is too expensive to use and support for such a centralized approach is prone to vulnerabilities.

This work proposes a decentralized approach which tries to address the problem of the single point of failure. To address this issue, the proposed work leverages the blockchain technology, which enables transparency, immutability integrity and openness of the data that we need to manage in a traceability application (Zheng *et al.*, 2016). This technology was introduced by Bitcoin and Satoshi Nakamoto with the paper "Bitcoin: A Peer-to-Peer Electronic Cash System" (Nakamoto, 2008) which explained a new open decentralized currency, backed up by cryptography, machine to machine interactions and consensus algorithm research, merged to produce a new and innovative technology.

A blockchain, can be explained as a ledger which keeps track of transactions happening on the decentralized network of bitcoin. Every user needs a public-private key-pair[2] which will act as the user's identity, and this makes the bitcoin network pseudo-anonymous (Koshy *et al.*, 2014). Bitcoin and blockchain technology in general, has been exploited in the past in many ways, to decentralize processes that was centralized until now. This mainly happened because of the ability to include metadata in the transactions happening. This can be achieved in the Bitcoin network, by leveraging a scripting language that is included in the network and the OP_RETURN[3] opcode in it. That way, data can be included in the immutable and decentralized ledger which can be trusted blindly due to the consensus model that the network is using.

---

[2] Public-key, or asymmetric, cryptography is an encryption scheme that uses two different keys. A private key which is known only by the owner of these keys, and the public key which results from the private key with a one-way algorithm, and can be known by anyone (Nechvatal, J., 1991).

The maximum amount of data that can be included in a Bitcoin transaction is 80 bytes (Bartoletti, 2017), but several transactions can be used in order to store more data. In many applications developed with blockchains, data do not get stored entirely, but only the hash of the needed data are stored, which give users the assurance that the data they are using are the actual data, by comparing the hash with the actual data, and that no tampering has happened.

In this paper we will present a method for achieving decentralization in a traceability application, which results in transparency, immutability, openness and security in the data that are crucial to a product's life cycle and are important to several users of these applications.

## 2. Example Application – Model Description

To demonstrate the decentralized traceability model, we will describe it by giving an example application which makes use of it. Our application, allows organizations and companies to trace the supply chain of a product in a decentralized and open way using the Bitcoin blockchain. The information that can be included in the blockchain may vary. From simple product information, to batch number and product serial number, the company will decide which information they will use with a client application. This information will become standard for this product and it will be used in order to update the information of the supply chain.

The platform's desirable characteristics are:

1. The company or organization generates an identity using the uPort[4] decentralized identity platform. This identity is unique, and it will be used to sign the information later.

2. An initialization transaction is generated so that it contains a hash of a text file produced by the client application. This file contains the list of information of a certain product type, that must be included in it when it gets updated in the blockchain.

3. From now on, companies will generate files that contain information about a certain batch or product, it will be signed by the uPort ID of the company, and the hash of the file will be included in the blockchain using an OP_RETURN transaction.

---

[3] Bitcoin uses a scripting language in order to construct transactions. It is a list of script words, called opcodes, and the OP_RETURN opcode is the standard way to include extra data in a
Bitcoin transaction. (Bartoletti, 2017)
[4] https://www.uport.me

4. To update the state of the product in a supply chain, companies will have to create a new file with the current state of the product and store it in the blockchain like earlier.

5. Revocation of a false state update has not been considered as a company can include that a previous state was false, in a new update transaction.

6. Files that have been generated will be stored in the Swarm[5] decentralized storage, with nodes created by the company of each product. That way, client users will be able to navigate through a web page and find information about a product supply chain state and compare the information file with the hash that was stored in the blockchain. By that, users will be able to trust the system about these information, and not the company, and the companies to be transparent about their product state information, in a decentralized and secure way.

7. Client users will use a client-side application that will automatically search for the file and the hash that are stored in the blockchain, download the file, compare it with the hash, and present the secure info to the user automatically. That way users will easily use the application, without any technical knowledge.

**2.1 Identity Creation**

Identities in our decentralized traceability application are managed by the uPort platform. The thing is that anyone can use this platform to create an identity, and this is not ideal, as anyone can impersonate someone else. For that reason, a digital signature with the company's information is mandatory to be included in the uPort ID, and an amount of verifications from other users must be included for a company, considering its size.

Client users can also have uPort IDs, to verify a company or report it, but it is not mandatory, because of the openness characteristic of the Bitcoin blockchain. That way users can simply use the application, or just download the Bitcoin blockchain, and check the stored information about a product, without any identity.

**2.2 Stored Data**

Data that are stored in a decentralized manner using blockchain technologies have to follow some guidelines. For our traceability model, a text file will be created using the XML language[6], in order to be able to set our own characteristics, using simple tagging, or even add new characteristics depending on the situation.

---

[5] Swarm is a distributed storage system based on the Ethereum blockchain. https://swarm-guide.readthedocs.io/en/latest/introduction.html

This file will need significant size availability by our system, which is a disadvantage of blockchains. As said, blockchains cannot store a huge amount of data, or even if they can, they are slow, and transaction fees will increase significantly. For that reason, a hash of the file is used, which is a fixed size string that can be produced by any amount of data provided to the cryptographic hash function. That way every time we need to store new information in a blockchain, the amount of data needed will be fixed and fees will be easily determined.

This file, and as a result the hash of the file, will be generated by a client application that will ask for a series of information from the user, at the initialization state, for the item to be stored in the blockchain. After the initialization step, state update files will be automatically created by the client application, which may relate to devices like barcode scanners, RFID receivers etc. These devices will give trace information to the application, which will store that information as an update, to the blockchain in use.

Also, text files will be stored in a decentralized manner with Swarm, as mentioned in Chapter 2. That way end users that need the stored information will be able to download the file and compare it with the hash that is stored in the Blockchain in use.

## 2.3 Transaction Coding Schema

For the client application to be able to differentiate the multiple situations that can happen in our system, a coding schema has to be developed which will accompany the hashes of the data files that are stored in the blockchain. The available codes that can be added before the hash, will be the following:

- "IT": Which concerns the Initialization Transaction of an item.
- "UT": Which concerns an Update Transaction with update information about the item.
- "RT": Which concerns a Revocation Transaction

**Figure 1:** The complete data model used to store the hash with the code schema in the blockchain.

| Code Schema | File Hash |
|---|---|

---

[6] Extensible Markup Language is a markup language that defines a set of rules for encoding documents in a format that is both human-readable and machine-readable.

As seen in Figure 1, in order to store the final hash, and for the client application to be able to detect correctly the stored data, a data store model has to be followed, where the code schema is in the beginning followed by the hash of the file. All this information is then added to the OP_RETURN field of the transaction that will store this data to the blockchain.

## 3. Discussion

As we can see, the process just generates an update to the current product state and stores it in the blockchain. The process is similar with what happens until now with traceability applications, but now we can exploit a decentralized technology and all its characteristics that were mentioned in Chapter 1.

Our example application is only theoretical, and a practical implementation has to be developed in order to find the true needs of this model. During our research the following improvements were discovered and can be considered in future development of our model.

- The bitcoin blockchain is mainly used for currency transactions, and already contains a huge amount of them. That means that the blockchain size is really bit, and to consider complete decentralization a user must download and store locally the whole blockchain. To address this issue, other blockchain implementation can be used, or even new private blockchains can be developed, with decreased amount of blockchain and transaction size.

- A model that will merge IDs has to be developed, for product owners, suppliers and other parts of the supply chain to be able to give information about the same product without the need of a different flow each time.

- Smart contracts[7] can be considered, in order to be able to store all the data to the blockchain, and avoid using files, signatures and hashes.

- Smart contracts or other blockchain implementations can be considered in order to reduce the transaction costs. Bitcoin has been a victim of currency trading, and the price is extremely volatile and high. This makes bitcoin only a theoretical choice, and other option must be examined.

- Bitcoin performance is poor. Transactions need a minimum of 10 minutes to get executed and a minimum of 30 minutes to be verified. This means that with the need of fast updates in a fast-paced supply chain, Bitcoin will fall sort and our model will fail. For that reason, several other blockchains can be considered, or even a new private blockchain with rules that will match the needs of the interested companies.

---

[7] Smart contracts in the Ethereum (Buterin, 2014) network (alternative blockchain), are self-executing contracts written in a Turing-complete programming language, that are stored decentralized in the blockchain.

As we can see, most of the improvements have to do with the use of the Bitcoin blockchain. But the use of it in our model is not accidental, as Bitcoin is the first to introduce the blockchain technology, and it is one of the most stable and old blockchains to date. Several other blockchains can be used, but they will not be as secure and as stable, with attacks like the 51% attack[8] and others being possible to target them.

## 4. Conclusions

While traditional traceability models can be effective, exploiting blockchain technologies can offer important improvements in traceability processes and applications. Several different implementations can be developed to match the client's needs, with private blockchains being the ideal solution if we have modularity and personalization in mind.

We have described a solution for this matter, using the Bitcoin blockchain to produce a decentralized traceability model, that will be open, secure, tamperproof and completely decentralized, without central authorities and entities.

While we focused on the traceability part of the supply chain management sector, blockchains and distributed ledger technologies can be applied in several other parts of it, to achieve decentralization or simply exploit some of the characteristics and advantages that come with the use of such technologies.

---

[8] An attack that can target a blockchains integrity. https://bitcoin.org/en/glossary/51-percent-attack